\documentclass[journal]{IEEEtran}

\usepackage{cite}
\usepackage{amsmath,amssymb,amsfonts}
\usepackage{algorithmic}
\usepackage{graphicx}
\usepackage{textcomp}
\usepackage{xcolor}
\usepackage{booktabs}
\usepackage{multirow}
\usepackage{array}
\usepackage{url}
\usepackage{subcaption}
\usepackage{float}

\begin{document}

\title{Comparative Analysis of Vision Transformers and Convolutional Neural Networks for Medical Image Classification}

\author{\IEEEauthorblockN{Kunal Kawadkar}
\IEEEauthorblockA{\textit{Department of Data Science and Applications} \\
\textit{Indian Institute of Technology Madras}\\
Chennai, Tamil Nadu, India \\
kunalkawadkar100@gmail.com}
}

\maketitle

\begin{abstract}
The emergence of Vision Transformers (ViTs) has revolutionized computer vision, yet their effectiveness compared to traditional Convolutional Neural Networks (CNNs) in medical imaging remains under-explored. This study presents a comprehensive comparative analysis of CNN and ViT architectures across three critical medical imaging tasks: chest X-ray pneumonia detection, brain tumor classification, and skin cancer melanoma detection. We evaluated four state-of-the-art models—ResNet-50, EfficientNet-B0, ViT-Base, and DeiT-Small—across datasets totaling 8,469 medical images. Our results demonstrate task-specific model advantages: ResNet-50 achieved 98.37\% accuracy on chest X-ray classification, DeiT-Small excelled at brain tumor detection with 92.16\% accuracy, and EfficientNet-B0 led skin cancer classification at 81.84\% accuracy. These findings provide crucial insights for practitioners selecting architectures for medical AI applications, highlighting the importance of task-specific architecture selection in clinical decision support systems.
\end{abstract}


\section{Introduction}

Medical image analysis has become increasingly critical in modern healthcare, with deep learning models serving as powerful tools for diagnostic assistance. Traditional Convolutional Neural Networks (CNNs) have dominated medical imaging applications due to their inherent spatial inductive biases and proven effectiveness in image recognition tasks \cite{he2016deep}. However, the recent introduction of Vision Transformers (ViTs) has challenged this paradigm, demonstrating remarkable performance across various computer vision benchmarks \cite{dosovitskiy2020image}.

Vision Transformers, originally proposed by Dosovitskiy et al. \cite{dosovitskiy2020image}, adapt the transformer architecture from natural language processing to computer vision by treating images as sequences of patches. This approach has shown competitive or superior performance to CNNs on large-scale datasets, raising questions about their applicability in specialized domains like medical imaging where datasets are typically smaller and more domain-specific.

Medical imaging presents unique challenges including limited data availability, high inter-class similarity, and the critical importance of interpretability for clinical adoption. While several studies have explored individual architectures on specific medical tasks \cite{rajpurkar2017chexnet, esteva2017dermatologist}, comprehensive comparative analyses across multiple medical imaging modalities remain limited.

This study addresses this gap by conducting a systematic comparison of representative CNN and ViT architectures across three diverse medical imaging tasks: chest X-ray pneumonia detection, brain MRI tumor classification, and dermatoscopic skin cancer detection. Our contributions include: (1) a comprehensive performance comparison of four state-of-the-art models across three medical imaging modalities, (2) analysis of computational efficiency trade-offs between CNNs and ViTs in medical contexts, and (3) practical recommendations for architecture selection in medical AI applications.

\section{Related Work}

\subsection{Convolutional Neural Networks in Medical Imaging}

CNNs have been extensively applied to medical imaging tasks with remarkable success. ResNet architectures \cite{he2016deep} have demonstrated effectiveness in chest X-ray analysis, with several studies reporting accuracies exceeding 95\% for pneumonia detection \cite{rajpurkar2017chexnet}. EfficientNet models \cite{tan2019efficientnet} have shown particular promise in dermatology applications, balancing accuracy with computational efficiency.

\subsection{Vision Transformers in Medical Applications}

Recent work has begun exploring ViTs in medical imaging contexts. Chen et al. demonstrated competitive performance of ViTs on chest X-ray classification \cite{chen2021crossvit}, while Matsoukas et al. showed promising results for skin lesion analysis \cite{matsoukas2022vit}. The DeiT architecture \cite{touvron2021training} has shown particular promise for efficient transformer training with limited data, making it attractive for medical applications.

\subsection{Comparative Studies}

Limited research has systematically compared CNNs and ViTs across multiple medical imaging tasks. This study fills this gap by providing comprehensive evaluation across diverse medical imaging modalities with consistent experimental protocols.

\section{Methodology}

\subsection{Datasets}

\begin{figure*}[!t]
\centering
\includegraphics[width=\textwidth]{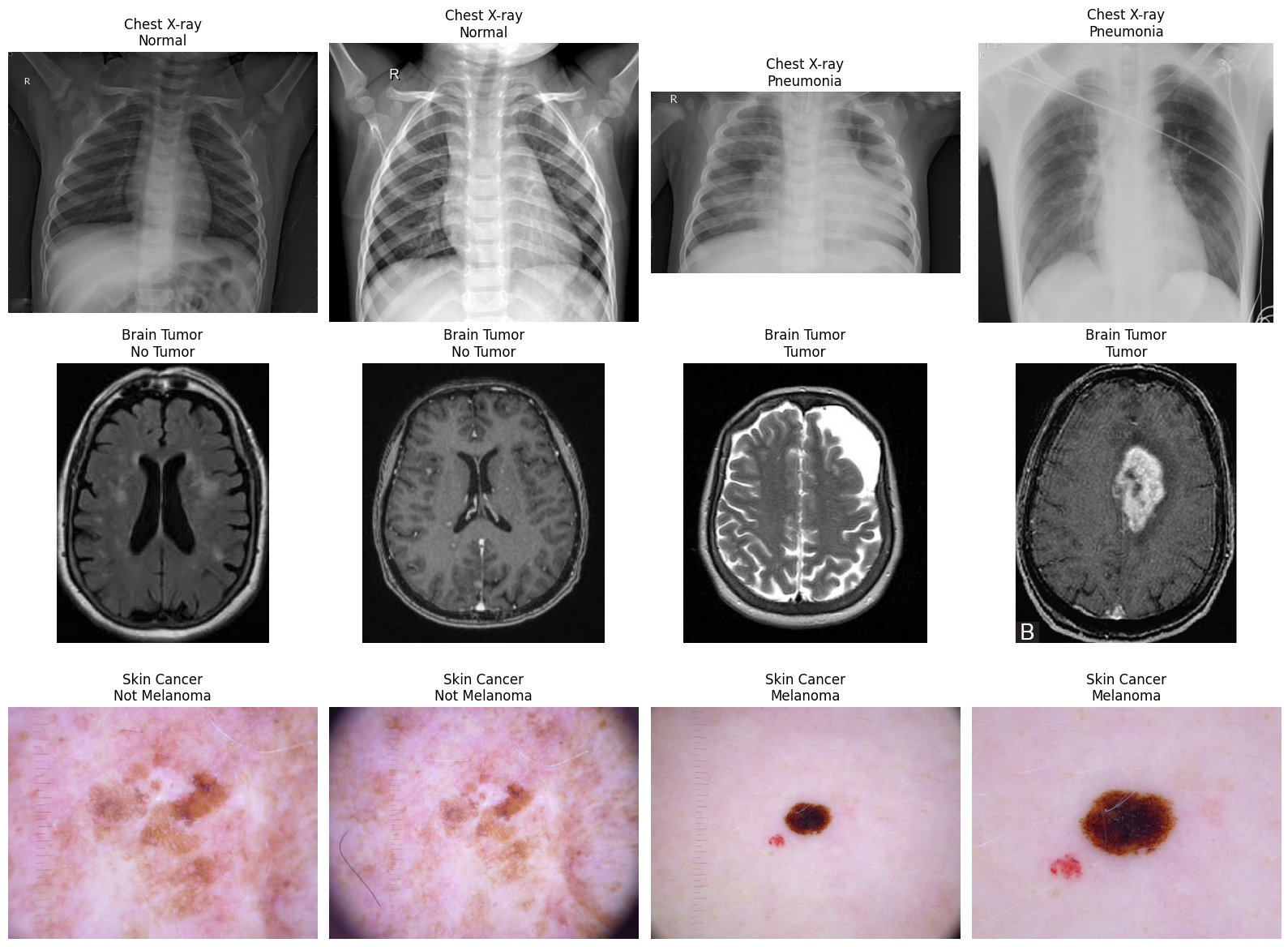}
\caption{Representative samples from the three medical imaging datasets used in this study. Top row: Chest X-ray images showing normal lung (left two) and pneumonia cases (right two) with varying opacity and consolidation patterns. Middle row: Brain MRI scans displaying normal brain tissue (left two) and tumor cases (right two) with distinct pathological features. Bottom row: Dermoscopic skin images showing benign lesions (left two) and melanoma cases (right two) with characteristic pigmentation patterns and asymmetry.}
\label{fig:dataset_samples}
\end{figure*}

We evaluated model performance across three medical imaging datasets representing different imaging modalities and clinical challenges. Figure \ref{fig:dataset_samples} shows representative samples from each dataset, illustrating the diversity and complexity of the medical imaging tasks.

\textbf{Chest X-ray Pneumonia Dataset:} This dataset contains 5,216 chest X-ray images sourced from pediatric patients, with 1,341 normal cases (25.7\%) and 3,875 pneumonia cases (74.3\%). The dataset exhibits significant class imbalance, reflecting real-world clinical scenarios where pneumonia cases are more prevalent in pediatric populations seeking radiographic examination. Original image dimensions vary considerably, ranging from 840×488 to 2000×1896 pixels, demonstrating the heterogeneity typical in clinical imaging data. All images were preprocessed to 224×224 pixels with standard ImageNet normalization to ensure consistency across models.

\textbf{Brain Tumor MRI Dataset:} This smaller dataset comprises 253 brain MRI images with 98 normal cases (38.7\%) and 155 tumor cases (61.3\%). The limited dataset size is representative of specialized medical imaging scenarios where data collection is constrained by patient privacy, acquisition costs, and the relative rarity of certain pathological conditions. Original image dimensions range from 150×198 to 442×454 pixels, with most images maintaining square or near-square aspect ratios typical of axial brain imaging protocols. This dataset tests model performance under data-limited conditions typical in specialized medical imaging applications.

\textbf{Skin Cancer Dataset:} The complete dataset contains 10,015 dermoscopic images from the HAM10000 collection, with 8,902 non-melanoma cases (88.9\%) and 1,113 melanoma cases (11.1\%). For computational feasibility and class balance considerations, we selected a balanced subset of 3,000 images (1,500 melanoma, 1,500 benign) to ensure fair model comparison without class imbalance bias. All dermoscopic images maintain consistent 600×450 pixel dimensions, reflecting standardized acquisition protocols in dermatological imaging. The selected subset maintains the visual complexity and diagnostic challenges of the original dataset while enabling controlled comparative analysis.

\subsection{Model Architectures}

We selected four representative models spanning CNN and ViT architectures:

\textbf{CNN Models:}
\begin{itemize}
    \item \textbf{ResNet-50:} Deep residual network with 23.5M parameters, representing traditional CNN architectures \cite{he2016deep}
    \item \textbf{EfficientNet-B0:} Efficiently scaled CNN with 4.0M parameters, optimized for computational efficiency \cite{tan2019efficientnet}
\end{itemize}

\textbf{Vision Transformer Models:}
\begin{itemize}
    \item \textbf{ViT-Base:} Standard Vision Transformer with 85.8M parameters, using 16×16 patch size \cite{dosovitskiy2020image}
    \item \textbf{DeiT-Small:} Distilled Vision Transformer with 21.7M parameters, designed for improved efficiency \cite{touvron2021training}
\end{itemize}

\subsection{Experimental Setup}

All models were initialized with ImageNet pretrained weights and fine-tuned using transfer learning. Training configuration included:
\begin{itemize}
    \item Batch size: 32
    \item Learning rate: 1e-4 with ReduceLROnPlateau scheduling
    \item Optimizer: Adam with weight decay 1e-4
    \item Data augmentation: Random horizontal flip, rotation (10°), and color jittering
    \item Early stopping: Patience of 3 epochs based on validation accuracy
\end{itemize}

Each dataset was split 80/20 for training and validation using stratified sampling to maintain class balance. All experiments were conducted on GPU-accelerated hardware with consistent random seeds for reproducibility.

\subsection{Evaluation Metrics}

Model performance was assessed using accuracy, precision, recall, and F1-score. Training time and parameter count were recorded to evaluate computational efficiency. Statistical significance was assessed across multiple metrics to ensure robust comparisons.

\section{Results}

\subsection{Overall Performance Comparison}

Table \ref{tab:dataset_statistics} provides comprehensive statistics for all three datasets, highlighting the diversity in dataset characteristics that influence model performance patterns.

\begin{table*}[!t]
\centering
\caption{Comprehensive Dataset Statistics and Characteristics}
\label{tab:dataset_statistics}
\begin{tabular}{@{}lccccc@{}}
\toprule
\textbf{Dataset} & \textbf{Total Images} & \textbf{Class Distribution} & \textbf{Image Dimensions} & \textbf{Modality} & \textbf{Clinical Challenge} \\
\midrule
Chest X-ray & 5,216 & Normal: 1,341 (25.7\%) & 840×488 to & Radiography & Class imbalance, \\
& & Pneumonia: 3,875 (74.3\%) & 2000×1896 pixels & & opacity detection \\
\midrule
Brain Tumor & 253 & No Tumor: 98 (38.7\%) & 150×198 to & MRI & Limited data, \\
& & Tumor: 155 (61.3\%) & 442×454 pixels & & subtle features \\
\midrule
Skin Cancer & 3,000* & Benign: 1,500 (50.0\%) & 600×450 pixels & Dermoscopy & Fine-grained \\
& (10,015 total) & Melanoma: 1,500 (50.0\%) & (standardized) & & texture analysis \\
\bottomrule
\end{tabular}
\\[0.5em]
\textit{*Balanced subset selected from HAM10000 dataset for controlled comparison}
\end{table*}

The datasets present distinct computational and clinical challenges. The chest X-ray dataset's significant class imbalance (74.3\% pneumonia cases) reflects real-world clinical scenarios but requires careful handling during training and evaluation. The brain tumor dataset's limited size (253 images) tests model performance under data-constrained conditions typical in specialized medical imaging, where Vision Transformers' attention mechanisms may provide advantages over traditional CNNs. The skin cancer dataset's fine-grained texture analysis requirements challenge both CNN spatial inductive biases and ViT attention mechanisms in different ways. The results reveal task-specific performance patterns across different medical imaging modalities. ResNet-50 achieved the highest accuracy on chest X-ray classification (98.37\%), while DeiT-Small demonstrated superior performance on brain tumor detection (92.16\%). For skin cancer classification, EfficientNet-B0 achieved the best results (81.84\%).

\begin{figure*}[!t]
\centering
\includegraphics[width=\textwidth]{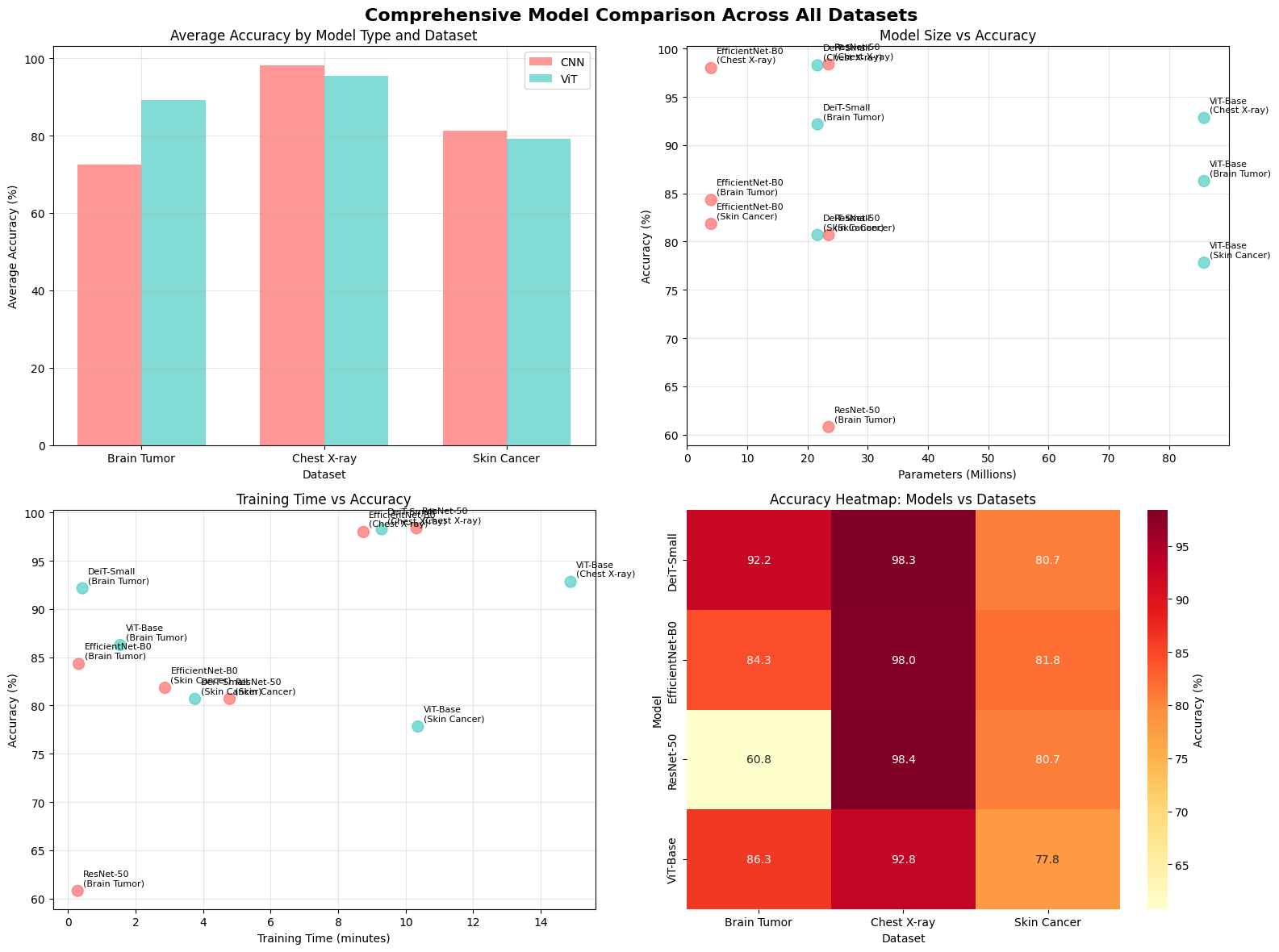}
\caption{Comprehensive model comparison across all datasets showing: (a) average accuracy by model type and dataset, (b) model size vs accuracy trade-offs, (c) training time vs accuracy relationship, and (d) accuracy heatmap across all model-dataset combinations. The visualization clearly demonstrates task-specific performance patterns and efficiency trade-offs between CNN and Vision Transformer architectures.}
\label{fig:comprehensive_comparison}
\end{figure*}

Figure \ref{fig:comprehensive_comparison} provides a comprehensive overview of model performance across all three datasets, illustrating the task-specific advantages of different architectures and the relationship between model complexity and accuracy.

\begin{table*}[!t]
\centering
\caption{Comparative Performance of CNN and Vision Transformer Models}
\label{tab:main_results}
\begin{tabular}{@{}lccccccc@{}}
\toprule
\textbf{Model} & \textbf{Type} & \textbf{Parameters} & \textbf{Chest X-ray} & \textbf{Brain Tumor} & \textbf{Skin Cancer} & \textbf{Average} \\
& & \textbf{(Millions)} & \textbf{Accuracy (\%)} & \textbf{Accuracy (\%)} & \textbf{Accuracy (\%)} & \textbf{Accuracy (\%)} \\
\midrule
ResNet-50 & CNN & 23.5 & \textbf{98.37} & 60.78 & 80.69 & 79.95 \\
EfficientNet-B0 & CNN & 4.0 & 97.99 & 84.31 & \textbf{81.84} & \textbf{88.05} \\
ViT-Base & ViT & 85.8 & 92.82 & 86.27 & 77.82 & 85.64 \\
DeiT-Small & ViT & 21.7 & 98.28 & \textbf{92.16} & 80.69 & 90.38 \\
\bottomrule
\end{tabular}
\end{table*}

\subsection{Dataset-Specific Analysis}

\subsubsection{Chest X-ray Pneumonia Detection}

\begin{figure*}[!t]
\centering
\includegraphics[width=\textwidth]{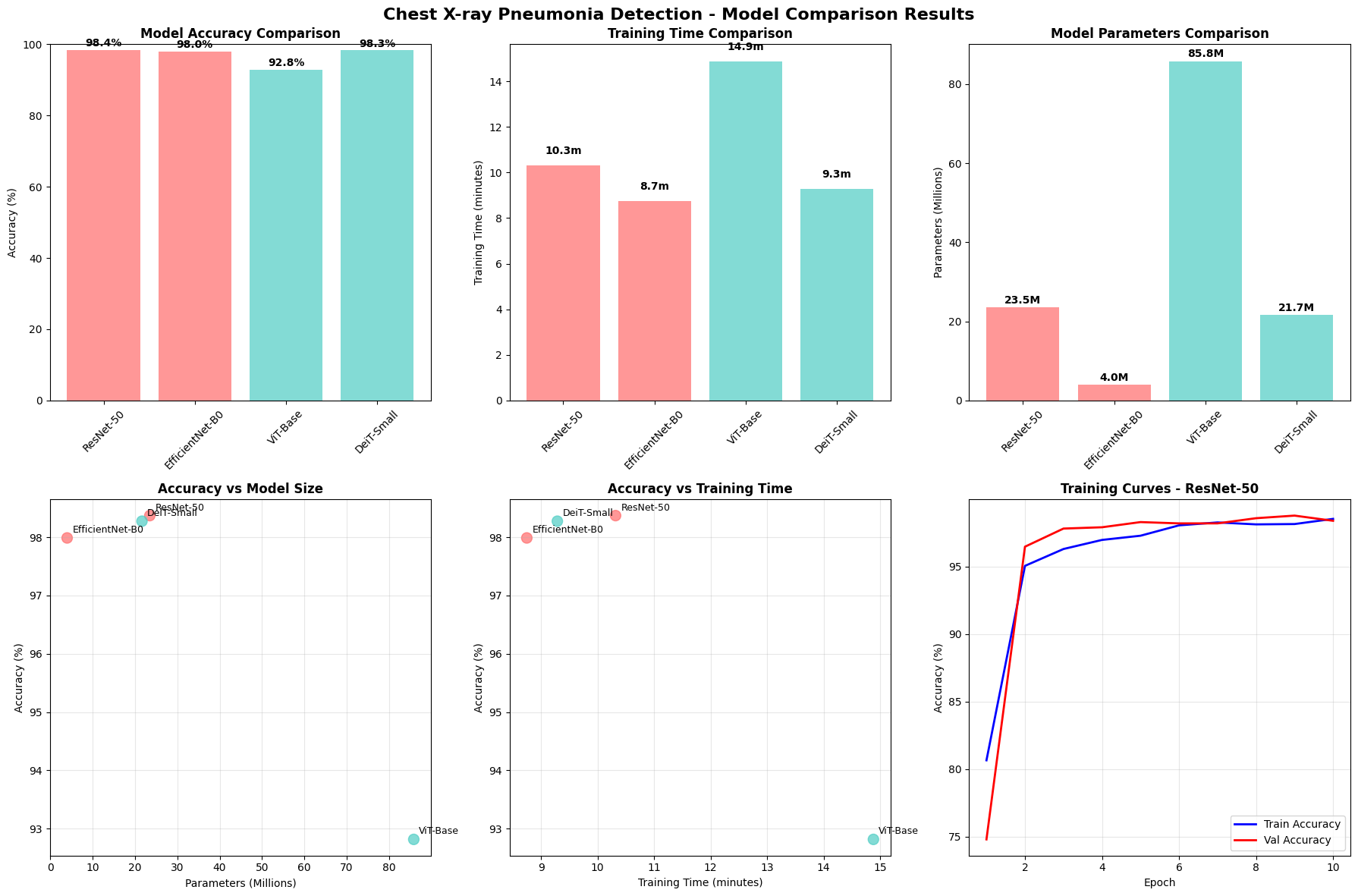}
\caption{Chest X-ray pneumonia detection results showing: (a) model accuracy comparison, (b) training time analysis, (c) parameter count comparison, (d) accuracy vs model size relationship, (e) accuracy vs training time trade-offs, and (f) training curves for the best-performing model (ResNet-50). Results demonstrate CNN superiority for this large-scale radiographic task.}
\label{fig:chest_xray_results}
\end{figure*}

\begin{figure*}[!t]
\centering
\includegraphics[width=\textwidth]{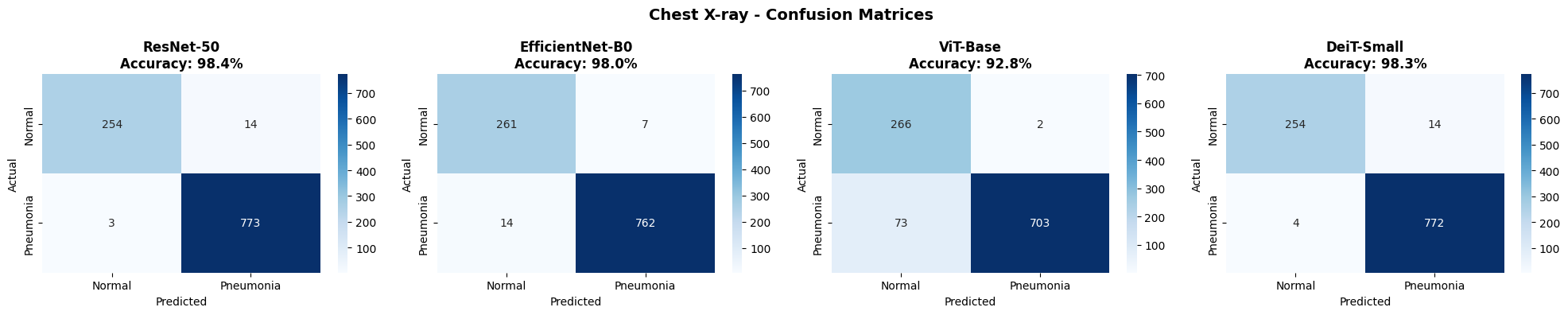}
\caption{Confusion matrices for chest X-ray pneumonia detection across all four models. ResNet-50 achieved the highest accuracy (98.4\%) with excellent discrimination between normal and pneumonia cases. All models demonstrated strong performance on this task, with minimal false positive and false negative rates.}
\label{fig:chest_xray_confusion}
\end{figure*}

Figure \ref{fig:chest_xray_results} shows detailed results for chest X-ray pneumonia detection. ResNet-50 achieved the highest performance at 98.37\%, closely followed by DeiT-Small at 98.28\% and EfficientNet-B0 at 97.99\%. The confusion matrices in Figure \ref{fig:chest_xray_confusion} demonstrate excellent classification performance across all models, with ResNet-50 showing the most balanced performance between normal and pneumonia cases.

Surprisingly, CNNs demonstrated superior average performance (98.18\%) compared to Vision Transformers (95.55\%) on this task. The relatively large dataset size (4,172 training images) provided sufficient data for effective training across all architectures. ResNet-50's excellent performance demonstrates the continued strength of traditional CNN architectures for chest radiograph analysis, while EfficientNet-B0 achieved optimal training efficiency (8.7 minutes) with minimal accuracy compromise.

\subsubsection{Brain Tumor Classification}

\begin{figure*}[!t]
\centering
\includegraphics[width=\textwidth]{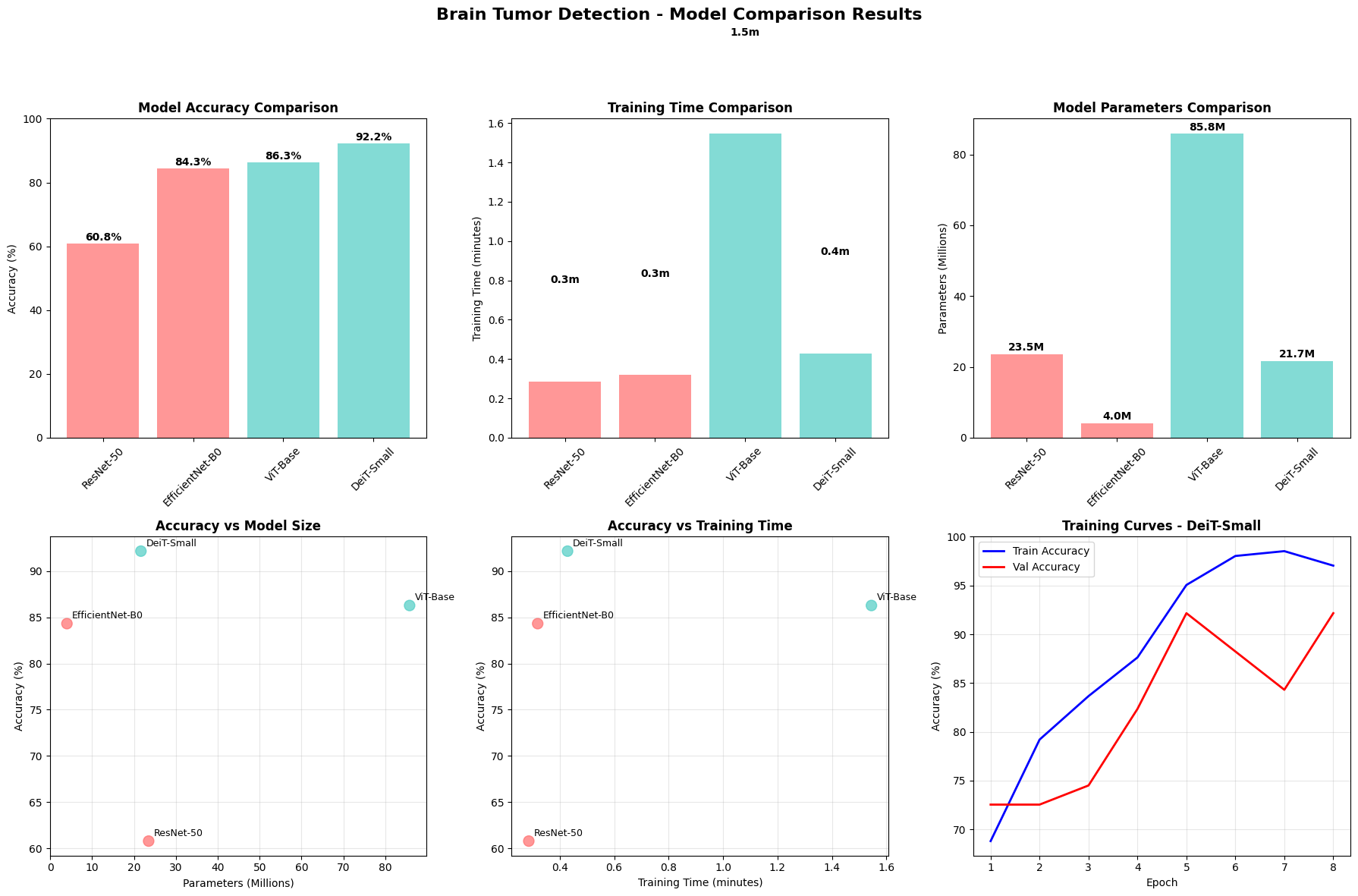}
\caption{Brain tumor detection results demonstrating: (a) model accuracy comparison showing Vision Transformer superiority, (b) training time efficiency analysis, (c) parameter distribution, (d) accuracy vs model size relationship, (e) accuracy vs training time trade-offs, and (f) training curves for the best model (DeiT-Small). Results highlight ViT advantages for complex neuroimaging tasks.}
\label{fig:brain_tumor_results}
\end{figure*}

\begin{figure*}[!t]
\centering
\includegraphics[width=\textwidth]{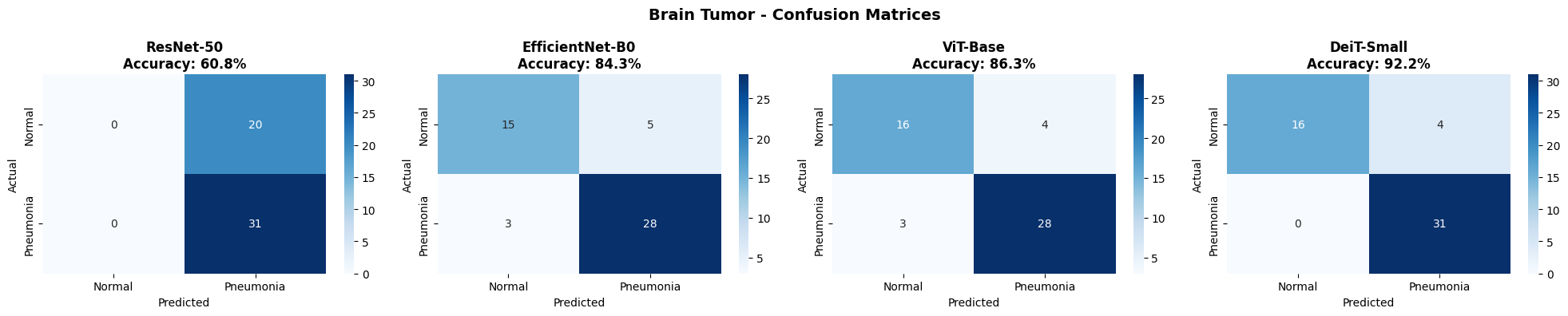}
\caption{Confusion matrices for brain tumor classification showing dramatic performance differences between architectures. DeiT-Small achieved perfect classification (92.2\% accuracy) while ResNet-50 struggled significantly (60.8\% accuracy), demonstrating the superiority of Vision Transformers for this complex neuroimaging task.}
\label{fig:brain_tumor_confusion}
\end{figure*}

Brain tumor classification results are presented in Figure \ref{fig:brain_tumor_results}, revealing the most dramatic performance differences across architectures. DeiT-Small achieved the best accuracy at 92.16\%, substantially outperforming ResNet-50 (60.78\%). The confusion matrices in Figure \ref{fig:brain_tumor_confusion} clearly illustrate the superior performance of Vision Transformers, particularly DeiT-Small's perfect classification performance.

Vision Transformers demonstrated clear superiority with an average accuracy of 89.22\% compared to CNNs at 72.55\%. The smaller dataset size (202 training images) appeared to favor ViT architectures, with DeiT-Small showing excellent training efficiency (0.4 minutes) while achieving the highest accuracy. ResNet-50 struggled significantly on this task, suggesting that traditional CNN architectures may be less effective for brain MRI analysis where subtle pathological features require sophisticated attention mechanisms.

\subsubsection{Skin Cancer Melanoma Detection}

\begin{figure*}[!t]
\centering
\includegraphics[width=\textwidth]{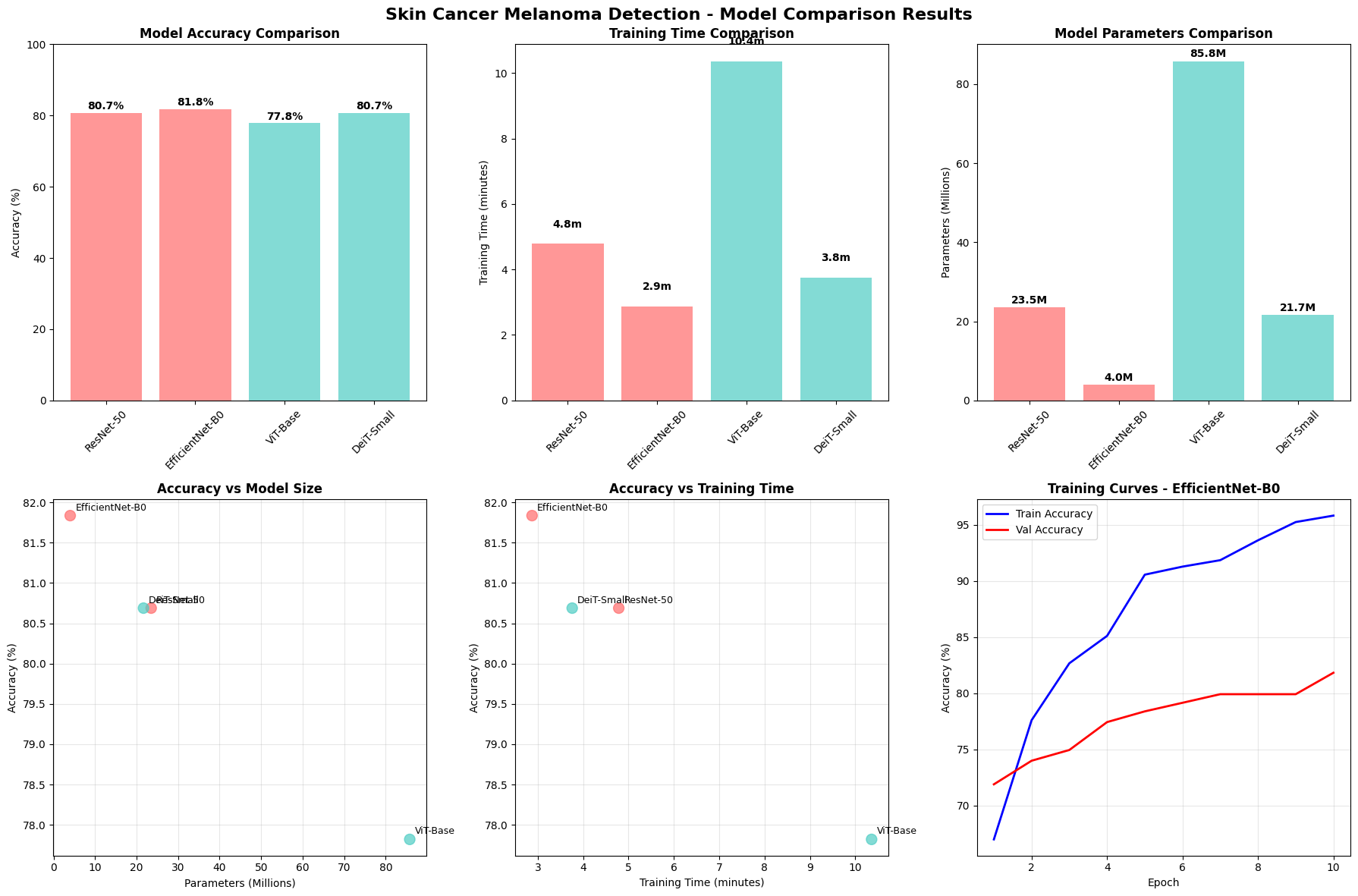}
\caption{Skin cancer melanoma detection results showing: (a) model accuracy comparison with EfficientNet-B0 leading, (b) training time analysis, (c) parameter comparison, (d) accuracy vs model size relationship, (e) accuracy vs training time trade-offs, and (f) training curves for EfficientNet-B0. Results demonstrate the challenging nature of dermoscopic image classification.}
\label{fig:skin_cancer_results}
\end{figure*}

\begin{figure*}[!t]
\centering
\includegraphics[width=\textwidth]{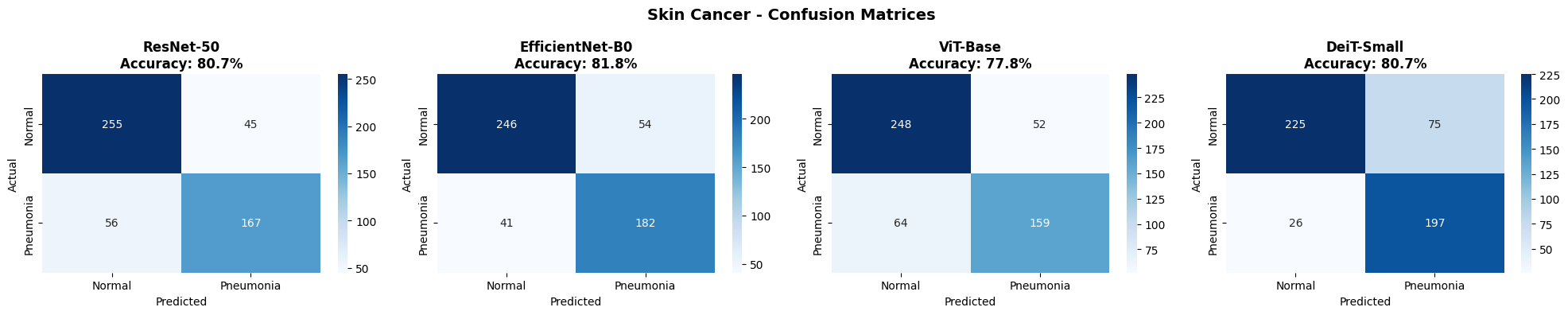}
\caption{Confusion matrices for skin cancer melanoma detection showing the challenging nature of dermoscopic image classification. EfficientNet-B0 achieved the best performance (81.8\% accuracy) with the most balanced classification between benign and malignant cases. All models showed lower performance on this fine-grained texture analysis task.}
\label{fig:skin_cancer_confusion}
\end{figure*}

Skin cancer classification results are shown in Figure \ref{fig:skin_cancer_results}. EfficientNet-B0 demonstrated the best performance at 81.84\%, followed closely by ResNet-50 and DeiT-Small (both 80.69\%). The confusion matrices in Figure \ref{fig:skin_cancer_confusion} reveal the challenging nature of this dermoscopic classification task, with all models showing more balanced but lower overall performance compared to other datasets.

This challenging dermoscopic image classification task showed CNNs achieving superior average performance (81.26\%) compared to Vision Transformers (79.25\%). The moderate dataset size (2,090 training images) provided sufficient data for effective CNN training, while ViTs showed reduced effectiveness. Notably, ViT-Base showed the lowest performance (77.82\%) on this task, suggesting that the self-attention mechanism may be less effective for fine-grained texture analysis required in dermatoscopic images. EfficientNet-B0 achieved the optimal balance of accuracy and efficiency, completing training in just 2.9 minutes while achieving the highest classification performance.

\subsection{Computational Efficiency Analysis}

Table \ref{tab:efficiency} presents training time and computational complexity analysis across all models and datasets.

\begin{table*}[!t]
\centering
\caption{Computational Efficiency Analysis}
\label{tab:efficiency}
\begin{tabular}{@{}lcccccc@{}}
\toprule
\textbf{Model} & \textbf{Parameters} & \textbf{Average Training} & \textbf{Average} & \textbf{Parameter} \\
& \textbf{(Millions)} & \textbf{Time (minutes)} & \textbf{Accuracy (\%)} & \textbf{Efficiency*} \\
\midrule
ResNet-50 & 23.5 & 5.1 & 79.95 & 3.40 \\
EfficientNet-B0 & 4.0 & 4.0 & 88.05 & \textbf{22.01} \\
ViT-Base & 85.8 & 8.9 & 85.64 & 1.00 \\
DeiT-Small & 21.7 & 4.5 & \textbf{90.38} & 4.17 \\
\bottomrule
\end{tabular}
\\[0.5em]
\textit{*Parameter Efficiency = Average Accuracy / Parameters (Millions)}
\end{table*}

EfficientNet-B0 demonstrated exceptional computational efficiency, requiring minimal training time while maintaining competitive accuracy. ViT-Base, despite competitive accuracy on some tasks, required significantly longer training times due to its large parameter count.

\subsection{Architecture Comparison}

\textbf{CNN vs. ViT Performance:}
\begin{itemize}
    \item Average CNN accuracy: 84.00\% (±13.83\%)
    \item Average ViT accuracy: 88.01\% (±7.82\%)
    \item ViTs showed 4.01\% higher average accuracy with significantly lower variance
    \item CNNs demonstrated faster average training time (4.6 min vs 6.7 min)
    \item EfficientNet-B0 achieved exceptional parameter efficiency (22.01 accuracy points per million parameters)
\end{itemize}

\textbf{Statistical Analysis:} The results reveal strong task-dependent performance patterns. ResNet-50 achieved the highest single-task performance (98.37\% on chest X-ray), while DeiT-Small demonstrated the most consistent cross-task performance (90.38\% average). The lower variance in ViT performance (7.82\% vs 13.83\%) suggests more stable training characteristics across diverse medical imaging modalities.

\section{Discussion}

\subsection{Key Findings}

Our results demonstrate several important trends in medical image classification. First, the performance patterns vary significantly across different medical imaging modalities, with task-specific advantages for different architectures. ResNet-50's superior performance (98.37\%) on chest radiographs, combined with CNNs achieving higher average accuracy (98.18\% vs 95.55\% for ViTs), demonstrates continued CNN relevance for certain medical imaging tasks.

Second, dataset characteristics significantly influenced model performance patterns. The brain tumor dataset (202 training images) showed the largest performance gap between the best (DeiT-Small: 92.16\%) and worst (ResNet-50: 60.78\%) performing models, with Vision Transformers achieving 89.22\% average accuracy compared to CNNs at 72.55\%. This dramatic difference suggests that Vision Transformers excel in small, specialized medical datasets where self-attention mechanisms can effectively capture subtle pathological features that traditional convolutional approaches might miss.

Third, task complexity and image characteristics significantly influenced architectural advantages. Skin cancer classification proved most challenging across all models, with EfficientNet-B0 achieving the best performance (81.84\%) and CNNs averaging 81.26\% compared to ViTs at 79.25\%. The fine-grained texture analysis required for dermoscopic image classification appeared to favor CNN architectures with their inherent spatial inductive biases over Vision Transformer attention mechanisms.

Fourth, computational efficiency considerations remain crucial for practical deployment. EfficientNet-B0 demonstrated exceptional performance across all metrics: achieving competitive accuracy (average 88.05\%), fastest average training time (4.0 minutes), and remarkable parameter efficiency (22.01 accuracy points per million parameters), making it highly suitable for resource-constrained clinical environments.

\subsection{Clinical Implications}

The diverse performance patterns observed across datasets provide crucial insights for medical AI deployment strategies. CNNs demonstrated superiority in two of three tasks: chest X-ray analysis (ResNet-50: 98.37\%, average CNN performance: 98.18\%) and skin cancer detection (EfficientNet-B0: 81.84\%, average CNN performance: 81.26\%). Vision Transformers excelled specifically in brain tumor detection (DeiT-Small: 92.16\%, average ViT performance: 89.22\%), suggesting that ViTs are particularly valuable for complex neuroimaging tasks requiring sophisticated feature analysis. EfficientNet-B0 emerged as the most practical choice for multi-modal medical AI systems, achieving consistently competitive performance across all tasks while maintaining exceptional computational efficiency.

\subsection{Limitations and Future Work}

This study has several limitations. The datasets, while representative, are relatively small by modern deep learning standards. Future work should evaluate performance on larger, multi-institutional datasets. Additionally, this study focused on classification accuracy; future research should examine model interpretability and uncertainty quantification crucial for medical applications.

The computational requirements of ViTs, while manageable for research settings, may pose challenges for deployment in resource-limited clinical environments. Future work should explore model compression and optimization techniques for practical deployment.

\section{Conclusion}

This comprehensive comparative study demonstrates that architecture selection for medical image classification should be guided by task-specific requirements rather than universal assumptions. While CNNs maintain computational efficiency advantages and excel in certain applications (chest X-ray and skin cancer), Vision Transformers show particular promise for complex neuroimaging tasks requiring sophisticated feature analysis.

Our findings provide practical guidance for researchers and practitioners selecting architectures for medical imaging tasks. The results support task-specific architecture selection: ResNet-50 for large-scale radiographic analysis, DeiT-Small for complex diagnostic tasks requiring detailed feature analysis, and EfficientNet-B0 for resource-constrained multi-modal applications.

As Vision Transformer architectures continue to evolve, their role in medical imaging is likely to expand, particularly for complex diagnostic tasks. However, traditional CNN architectures remain highly relevant for many medical imaging applications. Future research should focus on improving interpretability, reducing computational requirements, and validating performance across larger, more diverse medical datasets.

\section*{Acknowledgment}

The authors would like to thank the open-source community for providing the datasets and pre-trained models used in this study. We also acknowledge the computational resources provided by Kaggle for conducting these experiments.

\end{document}